\documentclass{mem}
\usepackage{natbib}\usepackage{txfonts}\usepackage{balance}
\usepackage{graphicx}
\usepackage[a4paper,breaklinks,dvipdfm]{hyperref}
\idline{000}{000}
\begin{document}

\def\dr{{\rm{d}}}
\def\gyr{{\rm Gyr}}
\def\henon{H\'enon}
\def\kpc{{\rm kpc}}
\def\msun{{\rm M}_{\odot}}
\def\myr{\rm Myr}
\def\rh{r_{\rm h}}
\def\tcr{\tau_{\rm cr}}
\def\tcrmax{\tau_{\rm cr1}}
\def\trh{\tau_{\rm rh}}
\def\rhoh{\rho_{\rm h}}
\def\rg{R_{\rm G}}

\title{The mass and radius evolution of globular clusters in tidal fields}

\author{Mark \,Gieles}
  \offprints{M. Gieles@surrey.ac.uk}
\institute{
Department of Physics -- University of Surrey, Guildford GU2 7XH , UK\\
\email{m.gieles@surrey.ac.uk}
}

\authorrunning{M. Gieles}

\titlerunning{The mass and radius evolution of globular clusters}

\abstract{We present a simple theory for the evolution of initially
  compact clusters in a tidal field. The fundamental ingredient of the
  model is that a cluster conducts a constant fraction of its own
  energy through the half-mass radius by two-body interactions every
  half-mass relaxation time. This energy is produced in a
  self-regulative way in the core by an (unspecified) energy source.
  We find that the half-mass radius increases during the first part
  (roughly half) of the evolution and decreases in the second half,
  while the escape rate is constant and set by the tidal field.  We
  present evolutionary tracks and isochrones for clusters in terms of
  cluster half-mass density, cluster mass and galacto-centric
  radius. We find substantial agreement between model isochrones and
  Milky Way globular cluster parameters, which suggests that there is
  a balance between the flow of energy and the central energy
  production for almost all globular clusters. We also find that the
  majority of the globular clusters are still expanding towards their
  tidal radius.  Finally, a fast code for cluster evolution is
  presented.  
\keywords{Galaxy: globular clusters -- Stars: kinematics and dynamics } 
} 

\maketitle{}

\section{Introduction}
Capturing cluster evolution in equations is complex because several
processes, including two-body relaxation, interactions with binary
stars, escape across the tidal boundary, and the internal evolution
and mass-loss of single and binary stars are all at work at the same
time. However, it is desirable to have a simple parameterisation of
the evolution of some fundamental cluster parameters such as mass and
radius \citep[e.g.][]{2008ApJ...689..919P}.

Here we provide a physically motivated and simple prescription for the
behaviour of the half-mass radius and tidal radius (i.e. mass). With
this we construct evolutionary tracks and isochrones for clusters
(i.e. not for the stars within them!) evolving in a tidal field.  This
forms the theoretical framework to explain empirically established
correlations between structural parameters and their environment as
found for Milky Way globular clusters
\citep[e.g.][]{1995ApJ...438L..29D,2000ApJ...539..618M} and for
extra-galactic globular cluster systems
\citep[e.g.][]{2005ApJ...634.1002J,2008MNRAS.384..563M,2010MNRAS.401.1965H}. We
do not aim to explain the shape and dependence on environment of the
globular cluster mass function.

The other principal exclusion is the evolution of the core parameters
(i.e. the core mass and radius).  Although the core is the place where
the energy is produced, the assumption of excluding it in the model is
justified by the discovery of \citet{1975IAUS...69..133H} that the
{\it rate of flow of energy is controlled by the system as a whole,
  and not by the core.}  In \henon's picture the mechanism of energy
generation in the core is self-regulatory and so we can assume that
the core produces the right amount of energy required by the system as
a whole.  This is comparable to the self-regulative energy production
in stellar cores, which was first realised by Eddington.  The
application of this idea to stellar dynamics was a breakthrough
allowing modellers to overcome the core collapse phase.
 
From $N$-body simulations of the long term (post-collapse) evolution
of single-mass clusters, i.e. where stars have the same mass, it was
found that binary stars act as the energy source
\citep{1994MNRAS.268..257G,2002MNRAS.336.1069B}.  In models with more
realistic initial conditions these binaries are usually considered to
be primordial.  Other mechanisms of energy generation have been
considered, including the action of a central intermediate-mass black
hole \citep{2004ApJ...613.1143B, 2007PASJ...59L..11H}

Mass-loss from stellar evolution can also provide the energy for the
dynamical evolution of clusters \citep{2010MNRAS.408L..16G}.
Typically most of the mass is lost from the most massive stars in the
cluster, that reside in the cluster core as the result of mass
segregation. The resulting energy production works together with
binaries in driving an expansion on relaxation time-scales
\citep{2010MNRAS.408L..16G}. Indeed, it may even dominate, as in the
specific example of 47 Tuc, a high-concentration cluster in which the
evolution of the core- and half-mass radii appear to be little
affected by the primordial binary population
\citep{2011MNRAS.410.2698G}.
                                   
These self-regulatory mechanisms of energy generation take time to
establish the balance between the energy generated in the core and the
energy requirements of the overall evolution of the cluster.  We refer
to the subsequent evolution as `balanced'.  In much research on
cluster dynamics this kind of evolution is usually associated with
`post-collapse' evolution, but this term is ambiguous; the phrase
`post-collapse' might be used for the entire evolution after the end
of mass segregation of massive stars ($\gtrsim10\,\msun$), whilst
others use it to refer to the evolution following the decrease in the
core radius after several Gyrs.  For this reason we prefer the term
`balanced evolution'.

In Section~\ref{sec:cycle} we present the model that unifies two
models of \henon. The evolution in the initial phase resembles that of
the model for the isolated cluster which expands with little loss of
stars \citep[][hereafter H61]{H61} and near final dissolution the
cluster resembles the `homologous cluster' \citep[][hereafter
  H65]{H65}, which contracts at a constant density.  Here the model is
described in its simplest form and we refer to \citep*[][hereafter
  G11]{G11} for more details.  In Section~\ref{sec:obs} the model is
compared to parameters of the Milky Way globular clusters. In
Section~\ref{sec:emacss} a fast code for cluster evolution based on
the theory presented in this work is presented.

\section{Model}
\label{sec:cycle}
Our attempt to unify the evolution of mass and radius of the two
models of \henon\ begins with one of the physical properties which the
two models have in common, i.e. a flux of energy at the half-mass
radius which is fed by an energy source in the core.  We constrain
ourselves to the energy flow at this radius, because we can then
construct a relatively simple set of relations for the behaviour of
the bulk properties of the cluster.  We adopt \henon's idealisation of
systems in which all stars have the same mass $m$.  We estimate the
total energy of the cluster as usual by

\begin{equation}
E =  -\alpha\frac{GM^2}{\rh},
\label{eq:E}
\label{eq:e}
\end{equation}
where $M$ is the mass of the cluster, $\rh$ is the half-mass radius,
and $\alpha\simeq0.2$ is a `form factor'.

In constructing a unified approximate model which includes the
transition from nearly isolated evolution to tidally limited
evolution, we assume that in both phases there is an energy flow due
to two-body relaxation of magnitude

\begin{equation}
  \frac{\dot E}{\vert E\vert} = \frac{\zeta}{\trh}.
  \label{eq:edote}
\end{equation}
Here $\trh$ is the half-mass relaxation time-scale and $\zeta$ is a
constant that can be interpreted as the efficiency of energy
conduction. In \henon's models $\zeta \simeq 0.08$\footnote{In G11 the
  values for the isolated and homologous cluster are derived from
  \henon's papers.} and from numerical simulations
\citet{2012MNRAS.422.3415A} find $\zeta\simeq0.1$.  We approximate the
expression for $\trh$ by assuming that the Coulomb logarithm is
constant, such that \citep{1987degc.book.....S}

\begin{equation}
\trh \propto N\tcr.
\label{eq:trh} 
\end{equation}
Here $N=M/m$ is the total number of stars and $\tcr$ is the crossing
time of stars in the cluster at the half-mass radius. We define $\tcr$
as
\begin{equation}
\tcr\equiv\left(G\rhoh\right)^{-1/2},
\label{eq:tcr}
\end{equation}
with $\rhoh\equiv3M/(8\pi\rh^3)$ the cluster density within the
half-mass radius.

Before proceeding further, we shall change the variables in which the
total energy $E$ is expressed, because this will facilitate the
further development of our model.  Instead of using the half-mass
radius, $\rh$, we shall use $\tcr$ such that $E\propto
-M^{5/3}\tcr^{-2/3}$.  Putting this together with our assumption about
the energy flux (equation~\ref{eq:edote}) we find

\begin{equation}
   -\frac{5}{3}\frac{\dot M}{M} +  \frac{2}{3}\frac{\dot\tcr}{\tcr} =  \frac{\zeta}{\trh}.
   \label{eq:flux-tidal}
\end{equation}
We recall that $\trh\propto M\tcr$ and thus
equation~(\ref{eq:flux-tidal}) has two variables: $M$ and $\tcr$. The
differential equation can be solved by relating $\dot M$ to $\tcr$.
Because $\dot M$ depends only on the orbit and is, to good
approximation, independent of the cluster mass and radius
\citep{1987ApJ...322..123L, 2008MNRAS.389L..28G} we can write for the
dimensionless escape rate $(\dot M/M)\trh = -(3/5)\zeta\tcr/\tcrmax$.
Here $\tcrmax$ is the maximum $\tcr$ which depends on the tidal field:
if the tides are weak, the cluster can expand to larger $\tcr$.
Combining this dimensionless escape rate with
equation~(\ref{eq:flux-tidal}) we find the dimensionless expansion
rate $(\dot\tcr/\tcr)\trh = (3/2)\zeta(1-\tcr/\tcrmax)$. Dividing
$\dot M$ by $\dot\tcr$ we find the surprisingly simple relation
between $M$ and $\tcr$

\begin{equation}
\frac{\dr M}{\dr \tcr} = \frac{2}{5}\frac{M}{\tcr-\tcrmax}.
\label{eq:dmdt}
\end{equation}
Integration gives an expression for $\tcr(M, \tcrmax)$, i.e. the
isochrones, and this, combined with equation~(\ref{eq:flux-tidal}),
can be used to get the time-dependent solutions $M(t)$ and $\tcr(t)$,
i.e. the evolutionary tracks. We do not give the functional forms
here, but instead refer the reader to G11.  In the next section we
proceed with a direct comparison between cluster isochrones and Milky
Way globular clusters parameters.

\section{Milky Way globular clusters}
\label{sec:obs}

\begin{figure*}
\center \includegraphics[width=14cm]{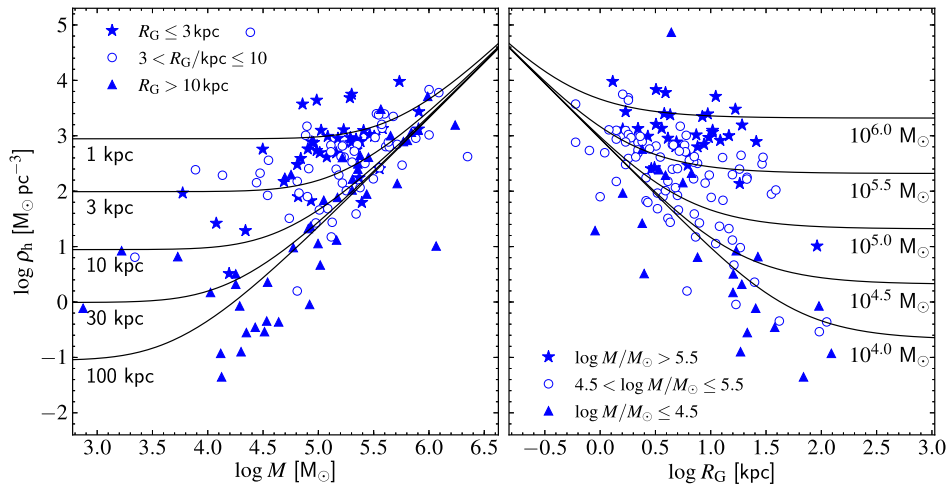}
     \caption{Isochrones based on the balanced evolution model of
       Section~\ref{sec:cycle}. In the left panel 5 isochrones for
       different $\rg$ values are shown and in the right panel 5
       isochrones for different $M$. Both panels show the 141 globular
       clusters in the \citet{1996AJ....112.1487H} catalogue from
       which $M$, $\rh$ and $\rg$ are available. In both
       panels different symbols are used for data in different $\rg$
       and $M$ regimes.}
   \label{fig:scaling}
\end{figure*}

\subsection{Are  clusters still expanding?}
\label{ssec:rel-imp}
In order to see if these types of predictions are relevant for real
globular clusters we compare our results to the globular clusters of
the Milky Way. We use the 2003 version of the
\citet{1996AJ....112.1487H} catalogue which contains entries for 150
globular clusters, and for 141 of them a luminosity, half-light radius
and galacto-centric radius determination are available.  To convert
luminosity to mass we adopt a mass-to-light ratio of 2
\citep{2005ApJS..161..304M} and we multiply the projected half-light
radius by 4/3 to correct for the effect of projection
\citep{1987degc.book.....S} and get an estimate for $\rh$.

The first thing we determine from the data is the fraction of globular
clusters that are in the expansion dominated phase.  We define the end
of the expansion phase as the moment where the time derivative of
$\trh$ is zero, which is when the fractional change in $M$ has the
same magnitude as the fractional change in $\tcr$, i.e. $\dr \ln M
/\dr\ln\tcr = -1$ (equation~\ref{eq:trh}). Combined with
equation~(\ref{eq:dmdt}) we then find that this happens when $M/M_0 =
(2/7)^{2/5} \simeq0.6$. In time this is when the cluster has evolved
for 40\% of its total life, because of the linear decrease of $M$.
The end of the expansion phase therefore depends on $M$ and the
mass-loss rate $\dot{M}$, which depends on the orbit.

We express $\dot{M}$ in terms of the galactocentric radius $\rg$.  We
assume that the Milky Way halo is an isothermal sphere and approximate
the $\rg$ dependent mass-loss rate by

\begin{equation}
\dot{M}\rg\simeq-20\,\msun\,\myr^{-1}\,\kpc,
\label{eq:mdot}
\end{equation}
which is in reasonable agreement with the evaporation rates found in
both $N$-body and Fokker-Planck models with a globular cluster type
stellar mass function (G11).  We adopt an age of $13\,\gyr$ for all
clusters.  Clusters that are now at the end of the expansion phase
have 60\% of their evolution, or $1.5\times13\,\gyr$, ahead of
them. The remaining life-time, or life expectancy, is defined as
$M/\dot{M}$ such that with equation~(\ref{eq:mdot}) we find that
clusters with a mass

\begin{equation}
  M\gtrsim 10^5\,\msun\left(\frac{4\,\kpc}{\rg}\right)\label{eq:mrg}
\end{equation}
\noindent
are still in the expansion-dominated phase.  This relation is
satisfied by 93 of the 141 clusters (i.e. roughly 2/3).  It follows
that the remaining 48 clusters (roughly 1/3) have expanded to the
tidal boundary and are in the evaporation-dominated phase.  This
perhaps surprising result has some interesting consequences. The most
important one is that the present day densities of the majority of the
globular clusters follow (roughly) from the self-similar expansion
model for isolated clusters: $\trh\propto M\tcr\simeq\,\,$constant,
i.e. $\rhoh\propto M^2$. This scaling relation is caused by internal
two-body relaxation and is independent of the tidal field; therefore a
similar scaling, with the same proportionality, should also hold for
extra-galactic clusters.  Moreover, in extra-galactic cluster samples
the fraction of clusters in the expansion-dominated phase is probably
larger; they are easier to detect because they have (on average)
higher mass (equation~\ref{eq:mrg}).

The prediction that a $\rhoh^{1/2}\propto M$ scaling must hold for the
majority of the Milky Way globular clusters is one of the main results
of this work.  

\subsection{Isochrones}
\label{ssec:isochrones}

Because all globular clusters have roughly the same age we focus on
isochrones with an age of $13\,\gyr$, rather than the evolutionary
tracks.  For a given age the isochrones can be expressed as
$\rhoh(M,\rg)$ and the detailed results are given in the appendices of
G11.  In Fig.~\ref{fig:scaling} we show the isochrones $\rhoh(M)$ for
several values of $\rg$ (left) and $\rhoh(\rg)$ for several values of
$M$ (right) diagrams together with the 141 globular clusters for which
data are available in the Harris catalogue.

In the left panel isochrones for clusters at different $\rg$ between
1\,kpc and 100\,kpc are shown. The isochrones roughly encompass the
data. The 100\,kpc isochrone clearly shows the asymptotic
$\rhoh^{1/2}\propto M$ behaviour following from expansion, which
roughly follows the lower envelope of data points. In the outer halo
the tidal field is so weak that all clusters with
$M\gtrsim10^4\,\msun$ are still expanding towards their tidal
boundary.  In the right panel the densities are shown as a function of
$\rg$ together with five isochrones for different masses. These
isochrones also roughly encompass the data.  The asymptotic behaviour
of the isochrones in both diagrams is given by labels in the two
diagrams.

\section{A fast code for cluster evolution}
\label{sec:emacss}
The model presented here makes several approximations in order to
facilitate simple analytical results. An improved version of the
model, which includes the escape of stars in the isolated regime, the
small $N$ dependence in the Coulomb logarithm and the delayed escape
of stars due to the anisotropic tidal field
\citep{2000MNRAS.318..753F} is presented in
\citet{2012MNRAS.422.3415A}. Here the differential equations for
$\dot{M}$ and $\dot{r}_{\rm h}$ are solved numerically with a
Runge-Kutta solver, which gives near instantaneous results for $M(t)$
and $\rh(t)$ as a function of the tidal field strength and the initial
cluster parameters which can be specified on the command line. The
code accurately reproduces the results of $N$-body integrations of
single-mass clusters \citep{2012MNRAS.422.3415A} and is publicly available on
\underline{https://github.com/emacss/emacss}. Future versions of the
code will be able to reproduce the evolution of more realistic stellar
clusters, including a stellar mass function and the effects of stellar
evolution.

\begin{acknowledgements}
MG thanks the organisers for organising a very pleasant conference and
Francesca D'Antona for her inspirational and enthusiastic contribution
to the 2012 Vatican Observatory Summer School!
\end{acknowledgements}

\bibliographystyle{aa}

\end{document}